\documentclass[%
aps,twocolumn,floatfix,showpacs,superscriptaddress,amsmath,amssymb]{revtex4-1}
\usepackage{graphicx}
\usepackage{dcolumn}
\usepackage{bm}
\graphicspath{ {./images/} }
\usepackage{color}
\usepackage{hyperref}
\usepackage{footnote}
\usepackage{wrapfig}
\usepackage{subcaption}

\usepackage{amsmath}
\usepackage{amssymb}
\usepackage{amsthm}
\usepackage{amsfonts}
\usepackage{braket}
\usepackage[T1]{fontenc}

 \usepackage{graphics}
 \usepackage{graphicx} 
 \usepackage{color,soul}
 \usepackage{textcomp}
 \usepackage{booktabs}
 \usepackage{tablefootnote}
 \usepackage{commath}
 \usepackage{physics}
 \usepackage{xfrac}

\begin{document}
\title{The effect of longitudinal spin-fluctuations on high temperature
properties of Co$_3$Mn$_2$Ge}

\author{Erna K. Delczeg-Czirjak}\email{erna.delczeg@physics.uu.se}
\affiliation{Department of Physics and Astronomy, Uppsala University,
Box 516, SE-75120 Uppsala, Sweden}
\author{O. Eriksson}
\affiliation{Department of Physics and Astronomy, Uppsala University, Box 516,
SE-75120 Uppsala, Sweden}
\affiliation{School of Science and Technology, \"Orebro University,
SE-$701 82$ \"Orebro, Sweden}
\author{A. V. Ruban}
\affiliation{Department of Materials Science and Engineering,
KTH Royal Institute of Technology, SE-100 44 Stockholm, Sweden}
\affiliation{Materials Center Leoben Forschung GmbH, A-8700 Leoben, Austria}

\begin{abstract}
 It is demonstrated that thermally induced longitudinal spin fluctuations (LSF) play an
 important role in itinerant Co$_3$Mn$_2$Ge at an elevated temperature.  The effect of
 LSF is taken into account during {\it ab initio} calculations via a simple model
 for the corresponding entropy contribution. We show that the magnetic entropy leads 
 to the appearance of a medium size local moment on Co atoms. As a consequence,
 this leads to a renormalization of the magnetic exchange interactions with a quite 
 substantial impact upon the caluclated Curie temperature. Taking LSF into account,
 the calculated Curie temperature can be brought to be in good agreement with
 the experimental value.
\end{abstract}

\date{\today}
\maketitle

\section{Introduction}

Recent studies show that high-throughput density functional theory (DFT) approach
\cite{AlenaHT, AlenaHT2, BATASHEV20211, screen3, screen4, screen5, screen6} can
be effective in filtering through a large number of compounds in the search for new
high-performance RE free permanent magnets or magnetocaloric materials, which would
be time-consuming and expensive to synthesize experimentally.
Curie temperature is one of the very important parameters for permanent magnets
and magnetocaloric materials filtering, which can be determined or at least 
qualitative estimated in ab initio calculations.

Within a DFT consideration, the Curie temperature is routinely obtained
using an additional statistical consideration based either on the energy difference
of magnetic structures or via magnetic exchange interactions of
a magnetic Hamiltonian. Although classical Heisenberg Hamiltonian
suffices to perform relatively accurate calculations of the Curie temperature,
the main problem in this approach may arise from the strong sensitivity of
the magnetic exchange interactions, for instance, as they are defined within
the magnetic force theorem \cite{Liechtenstein1984},  to the magnetic state
\cite{SGPM3}, which is also reflected in some cases in the strong dependence
of the local magnetic moment on the magnetic configuration \cite{Shallcross2005}.

This basically means that such a Hamiltonian cannot be used for such a
system within the whole range of temperatures and global magnetic states,
at least if one requires that its parameters should not depend on
a particular magnetic configuration, something which is the topic of recent
discussions \cite{SGPM3,cardias,dossantos}. Nevertheless, the whole formalism
of a classical Heisenberg Hamiltonian can still make sense and produce
reasonably accurate results when it is applied within a limited range 
of external parameters (temperature and pressure) and for a restricted
set of magnetic configurations. In this case, magnetic
exchange interactions should be determined at the corresponding conditions and
in the corresponding magnetic state.
In particular, since magnetic phase transitions are commonly of the
second order, one could argue that the corresponding magnetic exchange
interactions should be determined in the paramagnetic state, to which
such a transition happens. 

This is in fact the reason why the calculated Curie temperature of Co$_3$Mn$_2$Ge
\cite{AlenaHT} using magnetic exchange interactions obtained from the ordered,
ferromagnetic state, 750 K, was found to be twice as high as the
experimental one \cite{AlenaHT}, 359 K. Considering the fact that the
account of the experimentally observed chemical disorder between Co and
Ge did not improve the theoretical results, we assume here that the main
source of the discrepancy in \cite{AlenaHT} is the
strong dependence of magnetic interactions on the magnetic state.
 In other words, they are substantially different from the interactions in
paramagnetic state next to the point of magnetic phase transition.

The main reason for the quite strong dependency of the magnetic 
interactions in Co$_3$Mn$_2$Ge on the temperature and magnetic state is
its itinerant magnetism, especially related to Co atoms. In the usual
disordered local moment (DLM) ab initio zero K calculations \cite{DLM1,DLM2}
modeling paramagnetic state, the magnetic moment of Co atoms becomes too
small, while at finite temperature, its magnitude is strongly affected
by the LSF, which can be considered as thermal excitation due to 
specific entropy related to this degree of freedom
\cite{LSF1,LSF2,LSF3,LSF4,LSF_R2007,LSF_R2013}. The strong itinerant nature of
Co moments and the localization of Mn moments indicate a potential of 
Co$_3$Mn$_2$Ge as magneto caloric material similarly to Fe$_2$P based alloys
\cite{Fe2P_Erna,Dung2011a}. 

In this paper, we account for LSF in the paramagnetic state and calculate
the Curie temperature of Co$_3$Mn$_2$Ge using the corresponding magnetic
exchange interactions. The theoretical model and details of calculations
are described in the next section.


\section{Theoretical tools}

\subsection{Magnetic model}

A Heisenberg classical magnetic Hamiltonian is used for the corresponding
statistical thermodynamics simulations done using Monte Carlo (MC) method. 
In a general form, see for instance \cite{LSF_R2007}, it allows for
fluctuation of the magnitude of magnetic moments on each site of the
lattice, and thus it contains on-site terms as well as pair interaction
contribution, which depends on specific local magnetic moments at 
the corresponding sites and the average local magnetic moment of the
whole system. 

In the case of alloys, such an approach becomes quite cumbersome. 
Nevertheless,  it can be significantly simplified without loosing
much of the accuracy using its mean-field-like consideration 
for the task of finding the Curie temperature, which can be done
for the restricted range of temperatures. 

First of all, one can neglect the general temperature dependence of
the magnetic exchange interactions, connected with the temperature
dependence of the LSF, by considering one particular temperature,
which is close the experimentally known or "theoretically expected"
Curie temperature.
Secondly, although the fluctuation of local magnetic moments can be
quite large, some tests done for several pure metals (Ni, Co, and Fe)
show that the use of {\it average} local magnetic moments (for some
temperature and other external parameters) produces quite close results
to the "fluctuating" consideration. These allows one to simplify 
magnetic Hamiltonian to its usual form for alloys \cite{Ruban2017}:

\begin{equation} \label{eq:H_H_all}
H = - \sum_p \sum_{i,j \in p } \sum_{\alpha, \beta = \rm Co1, Co2, Mn}
J_p^{\alpha \beta} c_i^{\alpha}c_j^{\beta} \mathbf{e}_{i} \mathbf{e}_{j} .
\end{equation}
Here, $J_p^{\alpha \beta}$ are the magnetic exchange interactions
between $\alpha$ and $\beta$ alloy components for
coordination shell $p$ and $\mathbf{e}_{i}$ is the 
direction of the spin at site $i$;
$c_i^{\alpha}$ takes on value 1 if site $i$ is occupied by 
atom $\alpha$ and 0 otherwise.  

Statistical thermodynamics simulations of the magnetic phase transition 
were done by MC method implemented within the Uppsala  atomistic spin
dynamics (UppASD) software \cite{ASD1, ASD2}. MC simulations were performed
on a $40\times 40\times 40$ supercell with periodic boundary conditions.
The size and direction of the magnetic moments were chosen randomly at
each MC trial and 10000 MC steps were used for equilibration followed
then by 50000 steps for obtaining thermodynamic averages.


\subsection{Electronic structure and magnetic exchange interactions}

Electronic structure calculations were done by the exact muffin-tin orbital
(EMTO) method \cite{Andersen1994,Vitos2007} where the chemical and magnetic disorder
is treated within the coherent potential approximation (CPA) \cite{Soven1967, Gyorffy1972}
(EMTO-CPA \cite{Vitos2001}). The electrostatic correction to the single-site CPA
was considered as implemented in the Lyngby version of the EMTO code \cite{Ruban2016}.
For details the reader is referred to Refs. \cite{Ruban2016}, \cite{screen1}, and
\cite{screen2}.

The one-electron Kohn-Sham equations were solved within the soft-core and
scalar-relativistic approximations, with $l_{\rm max} = 3$ for partial waves
and $l_{\rm max}^{\rm t} = 5$ for their "tails". The Green's function was
calculated for 16 complex energy points distributed exponentially on a
semi-circular contour including states within 1.1 Ry below the Fermi level.
The exchange-correlation effects was described within the
local spin-density approximation \cite{LDA1, LDA2}. 
Magnetic exchange interactions were calculated within the magnetic force
theorem \cite{Liechtenstein1984} as is implemented in the Lyngby version 
of the EMTO-CPA code\cite{Ruban2016}. 

For "magnetic" alloy components, i.e. Co and Mn, the DLM configuration
was used in calculations within CPA. To account for the LSF, we
used the following approximation for the magnetic entropy \cite{Ruban2021}: 

\begin{equation} \label{eq_lsf1}
    S_{\rm mag} = d \ln(m) ,
\end{equation}    
where $m$ is local magnetic moment of an atom in the paramagnetic state,
$d=$ 1, 2, or 3, (case 1, 2, or 3 considered below) for the component
in high-, medium-, or low-spin states leading to different coupling between
longitudinal and transverse fluctuations of magnetic moment at finite
temperatures (at least above the magnetic transition). 

These expressions can be derived assuming a quadratic form for the LSF energy
with respect to the magnitude of the local magnetic moment following the recipe of
Ref. \cite{LSF_R2013}. The LSF energy can be determined in the DLM calculations.
Case 3 corresponds to the full coupling  between longitudinal and transverse
fluctuations, i.e. when local magnetic moment at finite temperature in the 
paramagnetic state can exists only due to LSF. In this case, the minimum 
of the LSF energy is at $m = 0$, like for pure Ni, \cite{LSF_R2007}.
Case 1 corresponds to a weak coupling, when longitudinal and
transverse fluctuations are little independent, and case 2 is an
intermediate case. 

\begin{figure}[h!]
  \centering
    \includegraphics[width=0.5\textwidth]{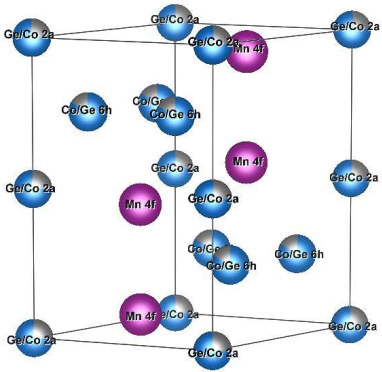}
    \caption{(Color online) Experimental structure of
    Co$_3$Mn$_2$Ge from Ref.~\cite{AlenaHT} generated by VESTA code \cite{VESTA}.}
    \label{str}
\end{figure}

All calculations were performed for the experimental structure and
composition given in Ref.~\cite{AlenaHT}. Co$_3$Mn$_2$Ge crystallizes in a
hexagonal structure with space group P63/mmc (number 194) with $a$ = 4.8032
$\AA$ and $c$ = 7.7378 $\AA$ lattice parameters, respectively. The actual
composition is Co$_{3.39}$Mn$_{2}$Ge$_{0.61}$ with a considerable intermixing
between Co and Ge, as follows: 84 at.~\% Co (labeled as Co1 on Fig.\ref{str}
and thereafter) and 16 at.~\% of Ge occupy the 6$h$ position, 74 at.~\% of Co
(labeled as Co2 on Fig.\ref{str} and thereafter) and 26 at.~\% Ge occupy the
2$a$ position, and Mn exclusively occupies the 4$f$ position.
Magnetic measurements show a non-collinear magnetic structure for 
Co$_{3.39}$Mn$_{2}$Ge$_{0.61}$ below 200~K and the ferromagnetic one
above that up to the Curie temperature at 359~K. 

\section{Results}

\begin{figure}[h!]
    \includegraphics[width=0.5\textwidth]{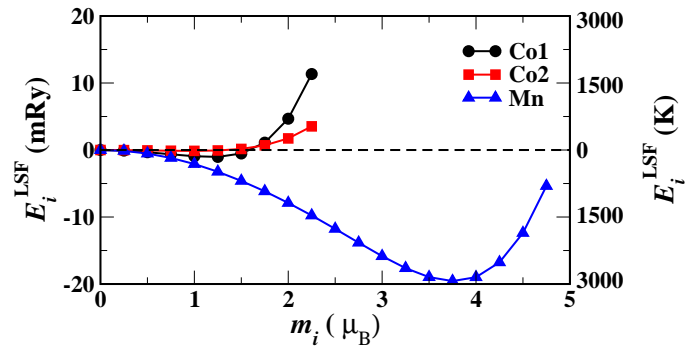}
    \caption{(Color online) Longitudinal spin fluctuation energy
    ($E^{\rm LSF}_i$) as a function of the local moment ($m_i$) of Co1 (black circles),
    Co2 (red squares), and Mn (blue triangles), respectively obtained in 
    zero K DLM calculations. The energy scale is given in mRy units
    (left) and absolute temperature (right).}
    \label{fig:Elsf1}
\end{figure}

To identify the type of magnetic behaviour of Co and Mn in paramagnetic
calculations, we calculate the LSF energy, $E_{\rm LSF} (m_i)$, in
the paramagnetic (DLM) state fixing the magnitude of the magnetic moment
of the corresponding alloy component while the others are relaxed to their
"equilibrium" magnitudes. It is shown in Fig.~\ref{fig:Elsf1} for Co1, Co2,
and Mn. Clearly, its behaviour differs substantially for Co and Mn:
in the case of Mn, it has a deep and pronounced minimum at $m_{\rm Mn} = $3.05
$\mu$B, which is very close to its magnitude in the FM ground state, 3.26 $\mu$B,
while the latter is quite shallow for both Co atoms. Obviously, Mn in this system is in
a more localized magnetic state, and it is not susceptible to LSF at finite
temperatures. This behaviour is in fact rather similar to that of the Fe$_2$P based magneto caloric materials \cite{Fe2P_Erna,Dung2011a}. Ref. \cite{Fe2P_Erna} shows that Fe moment on the tetrahedral 3$f$ site of Fe$_2$P is quite sensitive to the magnetic environment, while the Fe moment on the octahedral 3$g$ site is robust. This fact leads to a strong temperature dependence of the magnetic moments and at last is responsible for the first order nature of the magnetic transition and finally manifest in a large magnetocaloric effect in Fe$_2$P and related compounds \cite{Dung2011a}. 

At the same time, the T=0 K local moments of Co1 and Co2 are 0.44 and 0.12
$\mu$B, respectively, while their magnitudes in the FM ground state are
substantially higher: 1.60 and 1.57 $\mu$B. These means that LSF should affect
the magnetic moment of Co quite strongly, especially taking into consideration
the flat character of $E_{\rm LSF} (m_i)$ around the minimum.

To illustrate the effect of LSF at finite temperature on the magnitudes of local
magnetic moments of Co and Mn, we perform DLM-LSF calculations at the experimental
transition temperature 359 K using 3 different cases determined in the previous
section. The results are listed in Table~\ref{tab:mg}. As one can see, the magnetic
moments of Co change drastically with including LSF at even quite moderate temperature
in all three cases, although they are still lower in magnitude compared to their values of the FM
state. At the same time, the magnetic moment of Mn in the DLM-LSF calculations
at 359 K is practically the same as in DLM T=0 K calculations.

\begin{table}[th]
\caption{Element and site resolved magnetic moments ($m_i$ in $\mu_B$) and
theoretically estimated Curie temperatures ($T_{\rm C}$ in K) for
Co$_{3.39}$Mn$_{2}$Ge$_{0.61}$ in different magnetic states: FM, DLM,
and LSF-DLM for case 1, 2, and 3.}
\begin{tabular}{l|ccccc}
\hline \hline
$m_i$       & FM   &   DLM & LSF-1 & LSF-2 & LSF-3 \\
\hline
Co1         & 1.60 &  0.44 & 0.89  & 1.03   & 1.12 \\
Co2         & 1.57 &  0.12 & 0.86  & 1.02   & 1.11 \\
Mn          & 3.26 &  3.05 & 3.01  & 2.99   & 2.97 \\
\hline \hline
\end{tabular}
\label{tab:mg}
\end{table}

\begin{figure}
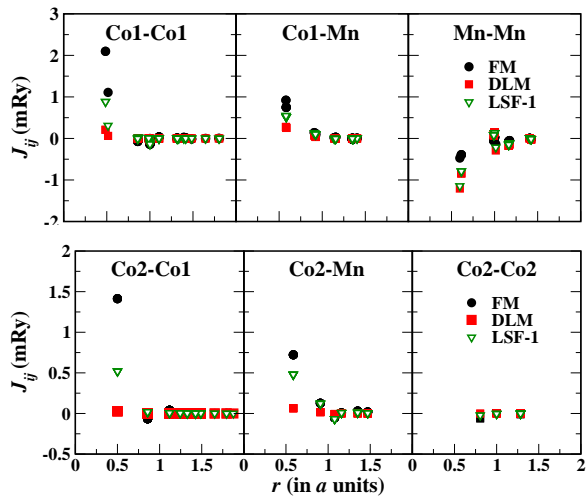

     \centering
     \begin{subfigure}[b]{0.45\textwidth}
         \centering
         \includegraphics[width=0.95\textwidth]{Jij1.eps}
     \end{subfigure}
     \vfill
     \begin{subfigure}[b]{0.45\textwidth}
         \centering
         \includegraphics[width=0.95\textwidth]{Jij2.eps}
    \end{subfigure}
      \caption{(Color online) Magnetic exchange interactions between the magnetic
      elements calculated for the three reference states. Black filled circles stand
      for FM, red filled squares denote DLM and open green triangles stand for case
      1 LSF at 359 K, respectively.}
        \label{fig:Jij}
\end{figure}

The effect of LSF on the local magnetic moments at finite temperature
translates to the values of the corresponding magnetic exchange interactions.
In Fig.~\ref{fig:Jij}, we present magnetic exchange interactions obtained
in different magnetic states: FM (black circles), DLM (red squares), and DLM-LSF1
(labelled LSF, green triangles) at 359 K, as a function of interatomic distance. As one can see, the strongest
interactions in the FM state are for between Co1-Co1 and Co1-Co2 atoms
at the first two coordination shells and they are of ferromagnetic type.
The Co$_i$-Mn interaction at the first coordination shell is just a bit smaller
than those interactions and it is also of the ferromagnetic type.
The Co2-Co2 magnetic exchange interactions are very small and they can
hardly influence the magnetic configuration at finite temperature.
The Mn-Mn interactions are also small and negative, i.e. it is of
antiferromagnetic type. Obviously, such interactions should strongly
stabilize the FM state at 0 K, which is not the ground state for this particular
alloy configuration.

The situation is quite different in the DLM state without LSF. All Co-Co
and Co-Mn interactions become insignificant,  while
the negative $J^{\rm MnMn}$ at the first several coordination shells
strengthens. Clearly, this type of interactions cannot provide the
stabilization of the FM state, and one can expect a stabilization either
of a certain type of antiferromagnetic or non-collinear state.
The DLM-LSF-1 magnetic exchange interactions at 359 K have somewhat
intermediate values for Co$_i$-Co$_i$ and Co$_i$-Mn, compared to those in
the FM and DLM states. At the same time, $J^{\rm Mn-Mn}$ are practically
the same as in the DLM state, which is quite expected.

\begin{figure}
    \includegraphics[width=0.47\textwidth]{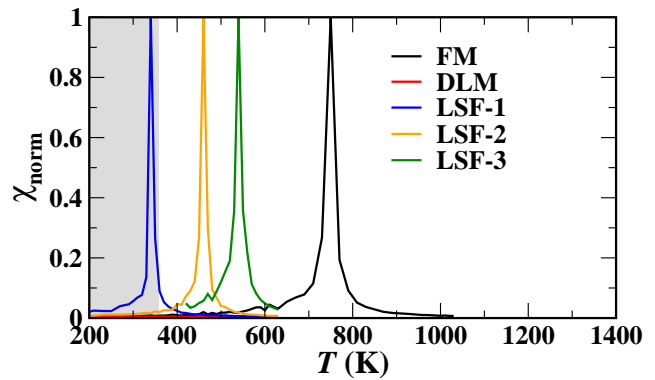}
    \caption{(Color online) Normalised magnetic susceptibility calculated for
    different reference states. Grey shaded area indicates the experimental
    ferromagnetic state.}
    \label{fig:susc}
\end{figure}

The magnetic exchange interactions obtained for different magnetic
states (also including LSF-2, and LSF-3 cases) were used in Monte Carlo
simulations of the magnetic phase transition. In Fig.~\ref{fig:susc},
we show the normalized magnetic susceptibility from these simulations, 
$\chi_{norm}$, for Co$_{3.39}$Mn$_{2}$Ge$_{0.61}$
obtained different sets of ($J_p^{\alpha \beta}$). 
The susceptibility peaks correspond to the magnetic transition at the
corresponding temperature, $T_{\rm C}$, which are also listed in
Table ~\ref{tab:T_c}. As is clear from Fig.~\ref{fig:susc}, the 
FM magnetic exchange interactions considerably overestimate
$T_{\rm C}$ compared to the experimental data (359~K) \cite{AlenaHT}.

\begin{table}[th]
\caption{Calculated Curie temperature for magnetic interactions
determined in different magnetic states.}
\begin{tabular}{l|ccccc}
\hline \hline
 State      & FM   &   DLM & LSF-1 & LSF-2 & LSF-3 \\
\hline
$T_{\rm C}$ &  750 &      -& 345   & 460    & 545\\
\hline \hline
\end{tabular}
\label{tab:T_c}
\end{table}

At the same time, DLM-LSF interactions produce quite good results,
especially in the DLM-LSF-1 case (DLM results are not shown since they
produce wrong ground state and very low transition temperature).
Although LSF-2 interactions produce higher Curie temperature than
experimental data, these results are still reasonable taking into
consideration that fact that the present simulations involve several 
practically unavoidable assumptions and approximations.

First of all, the model of LSF is quite rough: it is based on an
heuristic classical picture of magnetism within one-electron DFT at T=0 K.
Moreover, we do classical Heisenberg  Monte Carlo simulations, where
the LSF degree of freedom is  "hard-coded" in the corresponding effective
interactions. There are some approximations concerning the 
structure of the system too. Although we use experimental 
information about distributiuon of atoms between sublattices,
the latter is not completely certain since it is based on a fitting 
procedure involving some specific assumptions. We use
random distribution of Co and Ge atoms on their sublattices,
which just comes out from a random number generator in our MC
simulations. At the same time, Co and Ge atoms in real alloys
most probably have some specific atomic short range order.
Finally, our magnetic exchange interactions are determined at the
ideal, i.e. unrelaxed, lattice positions neglecting possible
local atomic relaxations related to the size mismatch of
Co and Ge. 

\section{Conclusions}

We have calculated the Curie temperature for disordered
Co$_3$Mn$_2$Ge in Monte Carlo simulations using magnetic exchange
interactions for pairs of Co and Mn atoms obtained in first principles
calculations for different magnetic states: FM, DLM, and DLM-LSF with
different degrees of coupling at 359 K. The FM interactions considerably
overestimate the transition temperature, while the DLM interactions are
too weak to produce reasonable results for magnetic transition.
The large difference between FM and DLM values of $T_{\rm C}$ is due
to the non-Heisenberg behavior of the system, i.e. the large difference between
the FM and the DLM local moments of the Co atoms. 

The failure of these two schemes is cured by account of thermally
induced longitudinal spin-fluctuations for Co atoms, which exhibit weak 
itinerant magnetism in this system. The effect of LSF is taken into
account during the {\it ab initio} calculations via a simple model that
includes the effect of thermally induced magnetic entropy on local moments and
consequently on $J^{\alpha \beta}$s. We show that the LSF contribution
is crucial for reconciliation of the theory and experimental data 
for the Curie temperature. This scheme is computationally very efficient
and easy to include to a high-throughput approach in searching new
candidates of permanent-magnet or magnetocaloric materials.

On the other hand, the strong dependence of the Co moments on the magnetic configuration, as a consequence on the temperature, and the stability of Mn moments,  indicate a promising magneto caloric potential of this material at room temperature similarly to Fe$_2$P based materials \cite{Fe2P_Erna, Dung2011a}.

\section*{Acknowledgements}
The authors thank the Swedish Foundation for Strategic Research (SSF),
project ''Magnetic materials for green energy technology'' (contract EM-16-0039)
for financing this project. STandUPP and eSSENCE are acknowledged for financial
support and the Swedish National Infrastructure for Computing (SNIC) for computational
resources (snic2021-1-36 and snic2021-5-340). O.E. also acknowledges support from
the Swedish Research Council (VR) and the Knut and Alice Wallenberg foundation (KAW).
Some DFT simulations were peformed on resources provided by the Swedish National
Infrastructure for Computing (SNIC) at PDC (Stockholm) and NSC (Linköping).
AVR acknowledges a European Research Council grant, the VINNEX center Hero-m,
financed by the Swedish Governmental Agency for Innovation Systems (VINNOVA),
Swedish industry,  and the Royal Institute of Technology (KTH).
AVR also gratefully acknowledges the financial
support under the scope of the COMET program within the K2 Center
“Integrated Computational Material, Process and Product Engineering (IC-MPPE)”
(Project No 859480). This program is supported by the Austrian Federal Ministries
for 718 Climate Action, Environment, Energy, Mobility, Innovation and Technology
(BMK) and for Digital and Economic Affairs (BMDW), represented by the Austrian
research funding association (FFG), and the federal states
of Styria, Upper Austria and Tyrol.
\bibliography{ref.bib}
\end{document}